\documentstyle[times,pramana,epsf,floats]{ias}
\begin{document}
\title{Hadrons in compact stars}

\author{Debades Bandyopadhyay}
\address{Saha Institute of Nuclear Physics, 1/AF Bidhannagar, Kolkata-700064,
India}
\pacs{2.0}

\abstract{ We discuss $\beta$-equilibrated and charge neutral 
matter involving hyperons and $\bar K$ condensates within relativistic models. 
It is observed that populations of baryons are strongly affected by the presence
of antikaon condensates. Also, the equation of state including $\bar K$ 
condensates becomes softer resulting in a smaller maximum mass neutron star.}

\maketitle
\section{Introduction}
There is a growing interplay between the physics of dense matter in 
relativistic heavy ion collisions and neutron stars \cite{Wal}.
Though Quantum Chromodynamics predicts a very rich
phase structure of dense matter, we can only probe a small region of it in the 
laboratories. Relativistic heavy ion experiments at CERN and BNL produce a 
hot (a few hundreds MeV) and dense matter (a few times normal nuclear matter
density) whereas the cold and dense matter relevant to neutron 
stars can not be produced in a laboratory. The study of dense matter in heavy
ion collisions reveals many new and interesting results such as the 
modifications of hadron properties in dense medium, the properties of strange
matter including hyperons and (anti)kaons and the formation of quark-gluon 
plasma. These
empirical informations from heavy ion collisions may be useful in understanding
dense matter in neutron star interior. On the other hand, 
satellite based observatories such as  Hubble space telescope and Chandra
X-ray observatory are pouring in very exciting data on compact stars. 
Measurements of masses and  radii of compact stars from various observations 
might constrain the composition and equation of state (EoS) of neutron star 
matter.
 
Neutron star matter encompasses a wide range of densities, from the
density of iron nucleus at the surface of the star to several times 
normal nuclear matter density in the core. The temperature of a neutron star
is a few MeV whereas the baryon chemical potential in its interior is a few
hundreds MeV. That is why neutron star matter is called the cold and dense 
matter. As chemical potentials of baryons and leptons increase with density
in the core, exotic forms of matter such as hyperons, Bose-Einstein condensate
of antikaons and quarks may appear there \cite{Gle}.

In this article, we discuss the composition and EoS of neutron star matter
involving Bose-Einstein condensates of antikaons within relativistic models.
Also, the structures of non-rotating neutron stars are calculated using this
EoS.

\section{Hadrons in cold and dense medium}
At normal nuclear matter density, neutron star matter mainly consists of 
neutrons, protons and electrons.  The particle population is so arranged as to 
attain a minimum energy configuration maintaining electrical charge neutrality 
and chemical equilibrium. At higher baryon density, hyperon formation becomes 
energetically favourable in neutron star interior as the total energy and 
pressure of the system are lowered by sharing baryon number among several 
baryon species. 

In compact star interior, hyperons maintain chemical equilibrium through weak 
processes. The generalised $\beta$-decay processes may be written in the form
$B_1 \longrightarrow B_2 + l+ \bar \nu_l$ and 
$B_2 +l \longrightarrow B_1 +\nu_l$ where $B_1$ and $B_2$ are baryons and 
l is a lepton. 
Therefore the generic equation for chemical equilibrium condition is
\begin{equation}
\mu_i = b_i \mu_n - q_i \mu_e ~,
\end{equation}
where $\mu_n$, $\mu_e$ and $\mu_i$ are respectively
the chemical potentials of neutrons, electrons and i-th baryon 
and $b_i$ and $q_i$ are baryon and electric charge of ith baryon respectively.
The above equation implies that there are two independent chemical
potentials $\mu_n$ and $\mu_e$ corresponding to two conserved charges i.e. 
baryon number and electric charge.

We adopt a relativistic field theoretical model to describe the pure hadronic
matter \cite{SD1}. The constituents 
of matter are ${n,p,\Lambda,\Sigma^+,\Sigma^-,\Sigma^0, \Xi^-,\Xi^0 }$ of the 
baryon octet and electrons and muons. In this model, 
baryon-baryon interaction is mediated by the exchange
of scalar and vector mesons and for hyperon-hyperon interaction, two 
additional hidden-strangeness mesons- scalar meson
$f_0$(975) (denoted hereafter as $\sigma^*$) and the vector meson $\phi$(1020) 
\cite{Mis} are incorporated.
Therefore the Lagrangian density for the pure hadronic phase is given by
\begin{eqnarray}
{\cal L}_B &=& \sum_B \bar\Psi_{B}\left(i\gamma_\mu{\partial^\mu} - m_B
+ g_{\sigma B} \sigma - g_{\omega B} \gamma_\mu \omega^\mu
- g_{\rho B}
\gamma_\mu{\mbox{\boldmath t}}_B \cdot
{\mbox{\boldmath $\rho$}}^\mu \right)\Psi_B\nonumber\\
&& + \frac{1}{2}\left( \partial_\mu \sigma\partial^\mu \sigma
- m_\sigma^2 \sigma^2\right) - U(\sigma) \nonumber\\
&& -\frac{1}{4} \omega_{\mu\nu}\omega^{\mu\nu}
+\frac{1}{2}m_\omega^2 \omega_\mu \omega^\mu
- \frac{1}{4}{\mbox {\boldmath $\rho$}}_{\mu\nu} \cdot
{\mbox {\boldmath $\rho$}}^{\mu\nu}
+ \frac{1}{2}m_\rho^2 {\mbox {\boldmath $\rho$}}_\mu \cdot
{\mbox {\boldmath $\rho$}}^\mu  + {\cal L}_{YY}~.
\end{eqnarray}
The isospin multiplets for baryons B $=$ N, $\Lambda$, $\Sigma$ and $\Xi$ are
represented by the Dirac spinor $\Psi_B$ with vacuum baryon mass $m_B$,
isospin operator ${\mbox {\boldmath t}}_B$ and $\omega_{\mu\nu}$ and 
$\rho_{\mu\nu}$ are field strength tensors. The scalar
self-interaction term \cite{Bog} is
\begin{equation}
U(\sigma) = \frac{1}{3} g_2 \sigma^3 + \frac{1}{4} g_3 \sigma^4 ~.
\end{equation}
The Lagrangian density for hyperon-hypron interaction (${\cal L}_{YY}$)
is given by,
\begin{eqnarray}
{\cal L}_{YY} &=& \sum_B \bar\Psi_{B}\left(
g_{\sigma^* B} \sigma^* - g_{\phi B} \gamma_\mu \phi^\mu
\right)\Psi_B\nonumber\\
&& + \frac{1}{2}\left( \partial_\mu \sigma^*\partial^\mu \sigma^*
- m_{\sigma^*}^2 \sigma^{*2}\right)
-\frac{1}{4} \phi_{\mu\nu}\phi^{\mu\nu}
+\frac{1}{2}m_\phi^2 \phi_\mu \phi^\mu~.
\label{Lag}
\end{eqnarray}

We perform this calculation in the mean field approximation \cite{Ser}. The 
mean values for
corresponding meson fields are denoted by $\sigma$ $\sigma^*$, $\omega_0$, 
$\rho_{03}$ and $\phi_0$. Therefore, we replace meson fields with their 
expectation values and meson field equations become,
\begin{equation}
m_{\sigma}^2\sigma=-{\partial U\over\partial\sigma}
+\sum_{B}g_{\sigma B}n_{B}^s~,
\end{equation}
\begin{equation}
m_{\sigma^*}^2\sigma^*=\sum_{B}g_{\sigma^* B}n_{B}^s~,
\end{equation}
\begin{equation}
m_{\omega}^2\omega_{0}=\sum_{B}g_{\omega B}n_{B}~,
\end{equation}
\begin{equation}
m_{\phi}^2\phi_{0}=\sum_{B}g_{\phi B}n_{B}~,
\end{equation}
\begin{equation}
m_{\rho}^2\rho_{03}=\sum_{B}g_{\rho B}I_{3B}n_{B}~.
\end{equation}
The scalar density and baryon number density 
\begin{eqnarray}
n_B^S &=& \frac{2J_B+1}{2\pi^2} \int_0^{k_{F_B}} 
\frac{m_B^*}{(k^2 + m_B^{* 2})^{1/2}} k^2 \ dk ~,
\end{eqnarray}
\begin{eqnarray}
n_B &=& (2J_B+1)\frac{k^3_{F_B}}{6\pi^2} ~, 
\end{eqnarray}
where Fermi momentum $k_{F_B}$, spin $J_B$, and isospin projection
$I_{3B}$. 

Effective mass and chemical potential of baryon $B$ are 
$m_B^*=m_B - g_{\sigma B}\sigma - g_{\sigma^* B}\sigma^*$ and
$\mu_{B} = (k^2_{F_{B}} + m_B^{* 2} )^{1/2} + g_{\omega B} \omega_0
+ g_{\phi B} \phi_0 + I_{3B} g_{\rho B} \rho_{03}$, respectively. 
In the pure hadronic phase, the total charge density is
\begin{equation}
Q^h = \sum_B q_B n^h_B -n_e -n_\mu =0~,
\end{equation}
where $n_B^h$ is the number density of baryon B in the pure hadronic phase and
$n_e$ and $ n_\mu$ are charge densities of electrons and muons respectively.

Solving the equations of motion in the mean field approximation 
along with effective baryon masses ($m_i^*$) and equilibrium 
conditions 
we immediately compute the equation of state in the pure hadronic phase.
The energy density ($\varepsilon^h$) is related to the pressure
($P^h$) in this phase through the Gibbs-Duhem relation 
\begin{equation}
P^h=\sum_i \mu_i n_i -\varepsilon^h~.
\end{equation}
Here $\mu_i$ and $n_i$ are chemical potential and number density for 
i-th species.

Kaplan and Nelson \cite{Kap} first showed in a chiral 
SU(3)$_L$ $\times$ SU(3)$_R$ model that Bose-Einstein condensation of $K^-$ 
mesons
could be possible in dense hadronic matter. They argued that the strongly 
attractive $K^-$-nucleon interaction might lower the effective mass and 
the in-medium energy of $K^-$ mesons in dense matter. The $s$-wave condensation sets in when the effective energy of $K^-$ mesons equals to its chemical 
potential. Recently, strongly attractive antikaon-nucleon interaction 
in-medium has been extracted from flow data of antikaons in heavy ion 
collisions \cite{Pal}.
  
In neutron star interior, strangeness changing processes such as,
\begin{equation}
N \rightleftharpoons N + \bar K,
\end{equation}
\begin{equation}
e^- \rightleftharpoons K^- + \nu_e.
\end{equation}
may occur. Here
$N\equiv (n,p)$ and $\bar K \equiv (K^-, \bar K^0)$ denote the 
isospin doublets for nucleons and antikaons, respectively.
The threshold conditions for $\bar K$ condensation
\begin{eqnarray}
\mu_{K^-} = \mu_n - \mu_p = \mu_e~,\nonumber\\
\mu_{\bar K^0} = 0~.
\end{eqnarray}
When the effective energy of $K^-$ mesons
($\omega_{K^-}$) equals to its chemical potential ($\mu_{K^-}$) which ,in turn,
is $\mu_e$, $K^-$ 
condensation sets in. Similarly, $\bar K^0$ condensation occurs when
$\omega_{\bar K^0} = \mu_{\bar K^0} = 0$.

The pure antikaon condensed phase is composed of baryons of the baryon octet, 
leptons and antikaons. In this phase, baryons are embedded in Bose-Einstein 
condensates.
The (anti)kaon-baryon interaction is treated on the same footing as 
baryon-baryon interaction. The Lagrangian density for (anti)kaons in the 
minimal coupling scheme \cite{SD1,GS98} 
\begin{eqnarray}
{\cal L}_K = D^*_\mu{\bar K} D^\mu K - m_K^{* 2} {\bar K} K ~,
\end{eqnarray}
where the covariant derivative 
\begin{eqnarray}
D_\mu = \partial_\mu + ig_{\omega K}{\omega_\mu} + ig_{\phi K}{\phi_\mu} 
+ i g_{\rho K} 
{\mbox{\boldmath $\tau$}}_K \cdot {\mbox{\boldmath $\rho$}}_\mu~. 
\end{eqnarray}
The isospin doublet for kaons
is denoted by $K\equiv (K^+, K^0)$ and that for antikaons is  
$\bar K\equiv (K^-, \bar K^0)$. The effective mass of (anti)kaons in this
minimal coupling scheme is given by
\begin{eqnarray}
m_K^* = m_K - g_{\sigma K} \sigma - g_{\sigma^* K} \sigma^* ~.
\end{eqnarray}
The equation of motion for kaons is given by,
\begin{equation}
\left(D_{\mu}D^{\mu} + m_K^{* 2} \right){K} = 0.~
\end {equation}
The s-wave ($\bf p=0$) dispersion relation for antikaons is,
\begin{equation}
\omega_{\bar K} = m_K^* - g_{\omega K} {\omega_0} - g_{\phi K} 
{\phi_0} + I_{3{\bar K}} g_{\rho K} {\rho_{03}}~, 
\end {equation}
where the isospin projection $I_{3K^-}=-1/2$ for $K^-$ and 
$I_{3{\bar K^0}}=1/2$ for ${\bar K^0}$.

The conserved current associated with (anti)kaons is derived by using
\begin{equation}
J_{\mu}^K = \left({\bar K}{\partial{\cal L}\over
{\partial^{\mu}{{\bar K}}}} - {\partial{\cal L}\over
{\partial^{\mu} K}}K\right)~.
\end{equation}
The density of antikaons is given by,
\begin{eqnarray}
n_{\bar K} &=& - J_{0}^K\nonumber\\
&=& 2 \left(\omega_{\bar K} + g_{\omega K} {\omega_0} + g_{\phi K} 
{\phi_0} - I_{3{\bar K}} g_{\rho K} {\rho_{03}}\right) K{\bar K}\nonumber\\ 
&=& 2 m_K^* K{\bar K}~. 
\end{eqnarray}
In the mean field approximation, the meson field equations in the presence 
of antikaons are, 
\begin{equation}
m_{\sigma}^2\sigma=-{\partial U\over\partial\sigma}+\sum_{B}g_{\sigma B}n_{B}^s
+ g_{\sigma K} \sum_{\bar K} n_{\bar K}~,
\end{equation}
\begin{equation}
m_{\sigma^*}^2\sigma^*=\sum_{B}g_{\sigma^* B}n_{B}^s
+ g_{\sigma^* K} \sum_{\bar K} n_{\bar K}~,
\end{equation}
\begin{equation}
m_{\omega}^2\omega_{0}=\sum_{B}g_{\omega B}n_{B}
- g_{\omega K} \sum_{\bar K} n_{\bar K}~,
\end{equation}
\begin{equation}
m_{\phi}^2\phi_{0}=\sum_{B}g_{\phi B}n_{B}
- g_{\phi K} \sum_{\bar K} n_{\bar K}~,
\end{equation}
\begin{equation}
m_{\rho}^2\rho_{03}=\sum_{B}g_{\rho B}I_{3B}n_{B}
+ g_{\rho K} \sum_{\bar K} I_{3 {\bar K}} n_{\bar K}~.
\end{equation}
The total charge density in the antikaon condensed phase is 
\begin{equation}
Q^{\bar K}=\sum_B q_B n_B^{\bar K} -n_K^--n_e -n_\mu =0
\end{equation}
where $n_B^{\bar K}$ is the baryon number density in antikaon condensed phase.

The total energy density in the $\bar K$ condensed phase consists of three 
terms, $\varepsilon = \varepsilon_B + \varepsilon_l + \varepsilon_{\bar K}$,
\begin{eqnarray}
{\varepsilon}
&=& \frac{1}{2}m_\sigma^2 \sigma^2 
+ \frac{1}{3} g_2 \sigma^3 + \frac{1}{4} g_3 \sigma^4
+ \frac{1}{2}m_{\sigma^*}^2 \sigma^{*2}\\ 
&& + \frac{1}{2} m_\omega^2 \omega_0^2 + \frac{1}{2} m_\phi^2 \phi_0^2 
+ \frac{1}{2} m_\rho^2 \rho_{03}^2 \\
&& + \sum_B \frac{2J_B+1}{2\pi^2} 
\int_0^{k_{F_B}} (k^2+m^{* 2}_B)^{1/2} k^2 \ dk\\
&& + \sum_l \frac{1}{\pi^2} \int_0^{K_{F_l}} (k^2+m^2_l)^{1/2} k^2 \ dk \\ 
&& + m^*_K \left( n_{K^-} + n_{\bar K^0} \right)~ .
\end{eqnarray}
And the pressure is given by,
\begin{eqnarray}
P &=& - \frac{1}{2}m_\sigma^2 \sigma^2 - \frac{1}{3} g_2 \sigma^3 
- \frac{1}{4} g_3 \sigma^4 \\
&& - \frac{1}{2}m_{\sigma^*}^2 \sigma^{*2} 
+ \frac{1}{2} m_\omega^2 \omega_0^2 + \frac{1}{2} m_\phi^2 \phi_0^2 
+ \frac{1}{2} m_\rho^2 \rho_{03}^2 \\
&& + \frac{1}{3}\sum_B \frac{2J_B+1}{2\pi^2} 
\int_0^{k_{F_B}} \frac{k^4 \ dk}{(k^2+m^{* 2}_B)^{1/2}}\\
&& + \frac{1}{3} \sum_l \frac{1}{\pi^2} 
\int_0^{K_{F_l}} \frac{k^4 \ dk}{(k^2+m^2_l)^{1/2}}~. 
\end {eqnarray}
The mixed phase of pure hadronic and antikaon condensed matter is described by
Gibbs phase equilibrium rules. Also, neutron star matter has two conserved
charges denoted by $\mu_n$ and $\mu_e$. Therefore,
The conditions of global charge neutrality and baryon number conservation are 
imposed through the relations \cite{Gle92}
\begin{equation}
(1-\chi) Q^h + \chi Q^{\bar K} = 0,\\
\end{equation}
\begin{equation}
n_B=(1-\chi) n_B^h + \chi n_B^{\bar K}~,
\end{equation}
where $\chi$ is the volume fraction of $K^-$ condensed phase in the mixed 
phase. The total energy density in the mixed phase is
\begin{equation}
\epsilon=(1-\chi)\epsilon^h + \chi \epsilon^{\bar K}~.
\end{equation}

\section{Composition and EoS of dense matter}
In the this model calculation, three distinct sets of coupling constants, 
meson-nucleon, meson-hyperon and meson-kaon, are required. 
Meson-nucleon coupling constants are generated by reproducing normal nuclear
matter properties - saturation density ($n_0$ = 0.153 fm$^{-3}$), binding 
energy, incompressibility and
symmetry energy. Here we exploit GM1 set for meson-nucleon coupling constants
\cite{SD1}. It is 
to be noted that nucleons do not couple with $\sigma^*$ and $\phi$.
 
Meson-hyperon coupling constants are determined from hypernuclei data and quark
model. The vector coupling constants for hyperons are obtained from SU(6)
symmetry as, 
\begin{eqnarray}
\frac{1}{2}g_{\omega \Lambda} = \frac{1}{2}g_{\omega \Sigma} 
= g_{\omega \Xi} = 
\frac{1}{3} g_{\omega N},\nonumber\\
\frac{1}{2}g_{\rho \Sigma} = g_{\rho \Xi} = g_{\rho N} ~~~ {\rm ;}~~~ 
g_{\rho \Lambda} = 0, \nonumber\\
2 g_{\phi \Lambda} = 2 g_{\phi \Sigma} = g_{\phi \Xi} = 
-\frac{2\sqrt{2}}{3} g_{\omega N}. ~ 
\end{eqnarray}
The scalar meson ($\sigma$) coupling to hyperons 
\begin{eqnarray}
U_Y^N(n_0) = - g_{\sigma Y} {\sigma} + g_{\omega Y} {\omega_0}~,
\end{eqnarray}
is determined from the knowledge of the potential depth of a hyperon in 
symmetric nuclear matter using hypernuclei data. For $\Lambda$ hyperon, this
potential depth is -30 MeV and it is -18 MeV for $\Xi$ hyperon. However, 
$\Sigma$ hyperon potential depth is +30 MeV. 

The $\sigma^*$-Y coupling constants are calculated by fitting them to a well
depth for a hyperon in hyperon (Y) matter at normal nuclear matter density,
${U_Y^{(Y^{'})}}(n_0)$, \cite{Gal} 
\begin{eqnarray}
U_{\Xi}^{(\Xi)}(n_0) = U_{\Lambda}^{(\Xi)}(n_0) = 2 U_{\Xi}^{(\Lambda)}(n_0) 
&=& 2 U_{\Lambda}^{(\Lambda)}(n_0)\nonumber\\
&& = -40~MeV.
\end{eqnarray} 

The scalar meson-kaon coupling constant is estimated from the real part of 
$K^-$ potential depth in normal nuclear matter density 
\begin{eqnarray}
U_{\bar K} \left(n_0\right) = - g_{\sigma K}\sigma - g_{\omega K}\omega_0 ~,
\end{eqnarray}
and the vector coupling constants from the quark model and isospin counting 
rule,
\begin{eqnarray}
g_{\omega K} = \frac{1}{3} g_{\omega N} ~~~~~ {\rm and} ~~~~~
g_{\rho K} = g_{\rho N} ~.
\end{eqnarray}
A strongly attractive antikaon potential depth of U$_{\bar K}$ = -160 MeV has
been used for this calculation. The strange meson
$\sigma^*$  and $\phi$ couplings with (anti)kaons are determined from 
the decay of f$_0$(975) and SU(6) symmetry relation respectively \cite{Mis}. 

The abundances of various species in $\beta$-equilibrated matter containing
baryons, electrons, muons and $K^-$ and $\bar K^0$ mesons are shown in 
Figure 1. We discuss the role of $K^-$ and $\bar K^0$ condensation on the
composition of $\beta$-equilibrated and charge neutral matter. 
\begin{figure}[htbp]
\epsfxsize=8cm
\centerline{\epsfbox{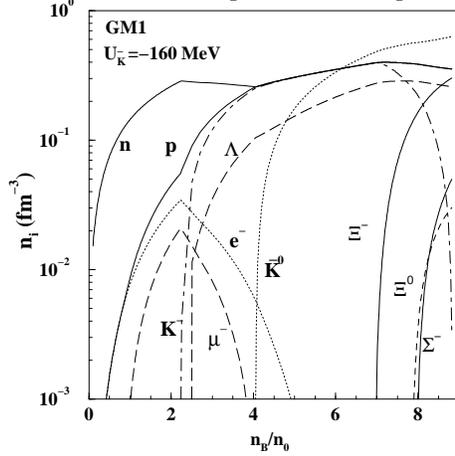}}
\vspace{-5cm}
\caption{The number densities n$_i$ of various particles in 
$\beta$-equilibrated hyperon matter including both $K^-$ and $\bar K^0$ 
condensates for GM1 set and antikaon optical potential depth at normal
nuclear matter density U$_{\bar K}$ = -160 MeV are plotted with baryon number
density.}
\label{fig:fig1}
\end{figure}
In the pure hadronic phase where local charge neutrality is imposed, 
abundances of 
nucleons, electrons and muons increase with density. Here, charge neutrality is
maintained among protons, electrons and muons. With the onset of $K^-$ 
condensation, the mixed phase begins at 2.23$n_0$. 
We find that $\Lambda$ hyperon is the first strange 
baryon to appear in the mixed phase at 2.51$n_0$. The total baryon density in 
the mixed phase is the sum of two contributions from hadronic and antikaon 
condensed phases weighted with appropriated volume fractions. As soon as $K^-$ 
condensate is formed, it rapidly grows with density and replaces electrons and 
muons. Being bosons, $K^-$ mesons in the lowest 
energy state are energetically
more favorable to maintain charge neutrality than any other negatively charged 
particles. Consequently, the proton density becomes equal to the density of
$K^-$ condensate. Also, the density of $\Lambda$ hyperon increases with baryon 
density in the mixed phase. On the other hand, the neutron density decreases in
the mixed phase. The reason behind it may be the creation of more protons in 
presence of $K^-$ condensate and also the growth of $\Lambda$ hyperons at the
expense of neutrons. The mixed phase terminates at 4.0n$_0$. 

Immediately after
the termination of the mixed phase, a second order $\bar K^0$ condensation sets
in $\sim$ 4.1$n_0$. 
With the appearance of $\bar K^0$ condensate, neutron and proton 
abundances become equal. The density of 
$\bar K^0$ condensate increases with baryon density uninterruptedly and even 
becomes larger than the density of  $K^-$ condensate. As soon as 
negatively charged hyperons - $\Xi^-$ and $\Sigma^-$ appear at higher 
densities, the density of $K^-$ condensate is observed to fall drastically. 
This is quite expected because it is energetically favorable for particles 
carrying  conserved baryon numbers to achieve charge neutrality in the system.
Leptons or mesons are no longer required for this sole purpose. Moreover,
lepton number or meson number is not conserved in the star. The system is 
dominated by $\bar K^0$ condensate in the high density regime.

The equation of state, pressure (P) versus energy density ($\varepsilon$)
for $\beta$-equilibrated and charge neutral matter with and without antikaon 
condensates are exhibited in Figure 2. 
\begin{figure}[htbp]
\epsfxsize=6cm
\centerline{\epsfbox{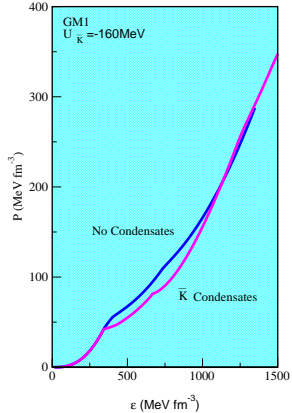}}
\vspace{-1.0cm}
\caption{The equation of state, pressure (P) versus energy density 
($\varepsilon$), for $\beta$-equilibrated hyperon matter with and without
$\bar K$ condensates.} 
\end{figure}
The lower solid line indicates
the overall EoS with hyperons and antikaon condensates whereas 
the EoS with hyperons and no condensate is exhibited by the upper solid line. 
Two kinks in the lower solid curve mark the beginning and end of the mixed 
phase where pure
hadronic and $K^-$ condensed phase are in thermodynamic equilibrium as 
dictated by Gibbs phase rules and global conservation laws. 
These kinks lead to discontinuity in the velocity of sound. 
The overall EoS with $\bar K$ condensates is softer compared with the EoS
without condensates. Consequently, the softer EoS gives rise to smaller maximum 
mass stars. The maximum mass of the static neutron star sequence calculated
with the EoS including $\bar K$ condensates is 1.57M$_{\odot}$ whereas that
of hyperon EoS without $\bar K$ condensates is 1.79M$_{\odot}$. 

\section{Summary}
In this article, we discuss the formation of hyperons and Bose-Einstein 
condensates of $\bar K$ mesons in cold and dense matter relevant
for neutron stars within relativistic models. Here we consider a first order 
$K^-$ condensation followed
by a second order $\bar K^0$ condensation. The populations of neutrons, protons
and hyperons are strongly modified in presence of both $K^-$ and $\bar K^0$ 
condensation. Also, antikaon condensates make the EoS softer compared with
the EoS without condensates and give rise to a smaller maximum mass neutron 
star.

\end{document}